# INCORPORATING MORPHOLOGICAL TYPES INTO SEMI-ANALYTIC SCHEMES FOR GALAXY FORMATION

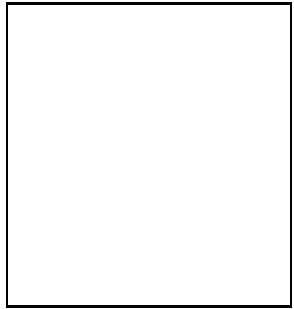

C.M. BAUGH[1]

[1] *Department of Physics, University of Durham, Science Laboratories, South Road, Durham DH1 3LE.*

**Abstract**

We test the hypothesis that elliptical galaxies are formed by violent mergers in a universe with hierarchical structure formation. Within the framework of a semi-analytic scheme for galaxy formation, we predict the distribution of morphological types with redshift and compare the colours of elliptical galaxies in different environments.

## 1 Introduction

The question of whether nature or nurture plays the biggest role in the formation of elliptical galaxies still remains unresolved. Did the gas in elliptical galaxies form stars on a shorter timescale than the collapse time of the halo? Alternatively, did the stars in disk galaxies get re-arranged into elliptical distributions during violent merger events?

Semi-analytic schemes have proved to be useful probes of the assumptions underlying the simple models used to represent the complex gravitational and gas dynamical processes in galaxy formation [1], [2], [3]. These schemes follow the star formation histories of dark matter halos of different sizes and are able to give reasonable first order matches to the observed galaxy luminosity function, counts and colours.

By extending the scheme of [3] to split the light of each galaxy into a bulge and disk component, we can test the hypothesis that elliptical galaxies are formed as the result of violent mergers by examining the distribution of morphological types and the colours of elliptical galaxies as a function of their environment.

## 2  The Galaxy Formation Scheme

Below we give a brief summary of the physical processes included in the galaxy formation scheme of [3].

The scheme follows the collapse and mergers of dark matter halos as a function of redshift. The gas inside each halo is shock heated to the virial temperature of the halo at collapse, and the fraction of gas that can cool over the lifetime of the halo, *i.e.* before the halo merges with another halo or before the output redshift, is calculated. The cold gas is then turned into stars giving rise to feedback processes which makes the star formation timescale and efficiency dependent upon the circular velocity of the dark matter halo in which the star formation is taking place.

When dark matter halos merge, the hot gas of the progenitor galaxies is stripped and is associated with the new halo. This gas can then cool onto the central galaxy of the halo, which is the most massive object in it. The satellite galaxies retain their stars and cold gas, and can only continue to form stars using their individual cold gas reservoirs. A dynamical friction timescale for the merger of the satellite galaxies with the central galaxy is calculated, with the galaxies merging if this time is shorter than the lifetime of the new dark matter halo. The form of the star formation feedback and galaxy merger parameterisations are motivated by the results of numerical simulations [4].

The galaxies are assigned luminosities in different bands using spectral energy distributions from the revised stellar population models of [5]. We neglect chemical enrichment in our model and compute luminosities assuming solar metallicity only. This means, for example, that we will not recover the same kind of colour magnitude relation that is observed for elliptical galaxies, which is believed to be driven by metallicity.

In this report, we shall use the parameters of the fiducial model of [3], with a standard Cold Dark Matter, $\Omega = 1$ universe.

## 3  Bulge Formation

We employ a similar scheme to that of [6] for the formation of galactic bulges. The galaxy formation models of [3] and [6], though in the same spirit, are different in detail. In particular, the star formation rate, feedback and galaxy merger timescale are all treated differently.

When stars form quiescently from gas that cools during the halo lifetime, the stars are added to the disk component of the galaxy. Hence, when a halo first collapses and forms stars, the stellar distribtuion will be a pure disk with no bulge component.

After a merger of dark matter halos, we compute the ratio of the mass of cold gas and stars in the accreted satellites to that in the central galaxy. If this ratio is larger than some specified parameter the merger is called a violent merger. In a violent merger, the disk of the central galaxy is destroyed. All the stars that are present are put into the bulge of the central galaxy. In addition, all the cold gas that is present is turned into stars in an instantaneous burst, with the new stars being added to the bulge.

If the merger is not classed as a violent merger, the disk of the central galaxy is preserved. The accreted stars are added to the bulge and there is no burst of star formation.

Irrespective of the type of merger event, gas that cools over the lifetime of the halo can form new stars that are then added to the disk of the central galaxy. In this scenario, the bulge to disk ratio, measured in terms of light or mass, is a dynamic quantity, determined as much by the quiescent evolution of the galaxy as by bursts of star formation.

## 4  Predictions

Kauffmann and collaborators [2], [6] have used the bulge to disk ratio in the B band to assign morphological types to their model galaxies. Based upon observed photometric decompositions [7], they define a spiral to be a galaxy in which less than 40% of the total light comes from the bulge component.

In Figure 1, we plot the fraction of galaxies that would be classed as spiral galaxies using the above criterion in halos of a given circular velocity. We have defined a violent merger to be one in which the central galaxy accretes more than 50% of its own mass. The points are coded to show the richness of the clusters and groups in terms of the number of galaxies brighter than $M_B - 5\log h = -18$. The upper panel shows the spiral fraction-circular velocity relation at redshift zero, the lower panel shows the results at redshift 0.5.

There is a large scatter in the spiral fraction recovered for clusters in our model that have intermediate circular velocities. Halos with $v_c > 800 \mathrm{kms}^{-1}$ (corresponding to a 1D velocity dispersion of $\sigma \sim 560 \mathrm{kms}^{-1}$ for an isothermal halo with an isotropic velocity distribution) do tend to have low spiral fractions, typically around 30%. At redshift 0.5, the richest clusters have a higher spiral fraction than at $z = 0$, reproducing the type of evolution in the membership of clusters seen by [8].

The elliptical galaxies in our models are the reddest, as expected in view of how the bulges are formed. We do however observe a difference in the colours of elliptical galaxies depending upon their environment. For ellipticals brighter than $V = -19$ found in halos with $v_c > 1000 \mathrm{kms}^{-1}$ we find a smaller scatter in the $B - V$ colours compared with ellipticals found in all halos. The cluster ellipticals are also systematically redder than the field ellipticals by $\Delta(B - V) \sim 0.04$, in agreement with the analysis of [9].

We shall compare the predictions of our model for the formation of galactic bulges against observations of galaxy counts and luminosity function evolution as a function of morphological type elsewhere [10].

## Acknowledgements

I would like to thank my collaborators on this project, Shaun Cole and Carlos Frenk. We are indebted to Stephan Charlot for providing us with a revised stellar population code ahead of publication.

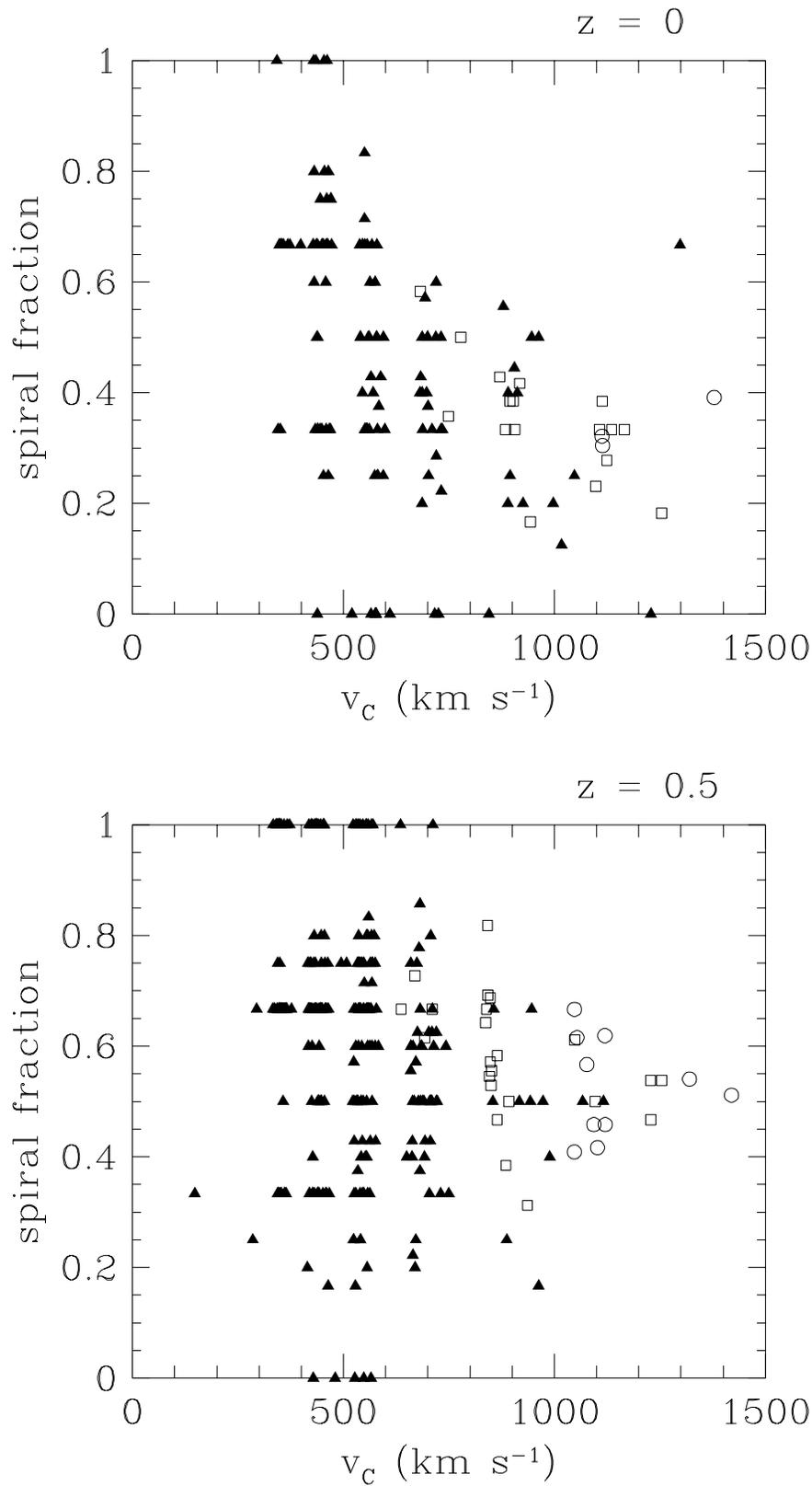

Figure 1: The spiral fraction of galaxies as a function of circular velocity. The symbols indicate the richness of the clusters and groups in terms of the number of galaxies brighter than $B = -18$. Triangles correspond to groups with between $2-10$ bright galaxies; squares have $10-20$ and circles have more than 20 bright members.